\documentclass[11pt]{amsart}
\usepackage[english]{babel}  
\usepackage[latin1]{inputenc}  
\usepackage{enumerate}  
\usepackage{amssymb,amsmath}  
\newcommand{\be}{\begin{equation}} \newcommand{\ee}{\end{equation}}
\newcommand{\bea}{\begin{eqnarray*}}
\newcommand{\eea}{\end{eqnarray*}} \newcommand{\beq}{\begin{eqnarray}}
\newcommand{\eeq}{\end{eqnarray}}
\newcommand{\RR}{\mathbb{R}}  
\newcommand{\NN}{\mathbb{N}}  
\newcommand{\CC}{\mathbb{C}}  
  
\newcommand{\EE}{\mathbb{E}}  
\newcommand{\PP}{\mathbb{P}}

\newcommand{\calA}{\mathcal{A}}  
\newcommand{\calB}{\mathcal{B}}
  
\newcommand{\calD}{\mathcal{D}}  
  
\newcommand{\calF}{\mathcal{F}}  
  
\newcommand{\calH}{\mathcal{H}}

\newcommand{\calL}{\mathcal{L}} 
  
\newcommand{\calN}{\mathcal{N}}

\renewcommand{\r}{\right}  
\renewcommand{\l}{\left}  
\newcommand{\la}{\langle}  
\newcommand{\ra}{\rangle}  
\newcounter{smalllist}  
  
\newcommand{\ko}{ k_{\omega}}  

\newcommand{\kD}{ k_\omega^D}  
\newcommand{\kDk}{ k_{H_\omega^{j}}}
  
\newcommand{\HD}{H_\omega^D}  
\newcommand{\HN}{ H_\omega}  
  
\newcommand{\ExpH}{\exp(-t H_\omega)}  
\newcommand{\ExpHD}{\exp(-t H_\omega^D)}  
\newcommand{\ExpHDk}{\exp(-t H_\omega^{j})}

\newcommand{\Erw}{ \mbox{\bf E}_x }  
\newcommand{\CcsmoothX}{C_c^{\infty} (X)}  
\newcommand{\CcsmoothD}{C_c^{\infty} (D)}  

\newcommand{\vol}{{\mathrm{vol}}}

\newcommand{\tr}{\mathop{\mathrm{tr}}}  
\newcommand{\esssup}{\mathop{\mathrm{ess\, sup}}}

\newtheorem{theorem}{Theorem}  
\newtheorem{lemma}{Lemma}[section]  
\newtheorem{prop}[lemma]{Proposition}  
\newtheorem{coro}[lemma]{Corollary}  
\theoremstyle{definition}  
\newtheorem{definition}[lemma]{Definition}  
  
\theoremstyle{remark}  
\newtheorem{remark}[lemma]{Remark}

\begin{document}  
\title[Integrated density of states for random metrics]{Integrated
density of states\\ for random metrics on manifolds}
\author[D.~Lenz]{Daniel Lenz} \author[N.~Peyerimhoff]{Norbert
Peyerimhoff} \author[I.~Veseli\'c]{Ivan Veseli\'c}
\address[D.~Lenz]{Fakult\"at f\"ur Mathematik, TU Chemnitz, 09107
Chemnitz, Germany} \email{dlenz@mathematik.tu-chemnitz.de}
\urladdr{www.tu-chemnitz.de/mathematik/analysis/dlenz}
\address[N.~Peyerimhoff]{Fakult\"at f\"ur
Mathematik\\Ruhr-Universit\"at Bochum, Germany}
\email{peyerim@math.ruhr-uni-bochum.de}
\urladdr{www.ruhr-uni-bochum.de/mathematik10/Norbert.html}
\address[I.~Veseli\'c]{Postdoctoral research fellow of the Deutsche Forschungsgemeinschaft, {\normalfont visiting the} Department of  Mathematics 253-37, California Institute of Technology, CA 91125, USA} 
\email{veselic@caltech.edu}
\urladdr{www.its.caltech.edu/\protect{\symbol{20}}veselic}
  
\date{\today}  
  
  
\begin{abstract}  
We study ergodic random Schr\"odinger operators on a covering
manifold, where the randomness enters both via the potential and the
metric.  We prove measurability of the random operators, almost sure
constancy of their spectral properties, the existence of a
selfaveraging integrated density of states and a \v{S}ubin type trace
formula.
\end{abstract}  
\maketitle

\section{Introduction}  
  
The mathematically rigorous study of random Schr\"odinger  
operators commenced in the  70ties. The motivation was to  
understand the transport properties of random media. Since then a  
variety of results on the spectral, wave-spreading and conductance  
properties of random Schr\"odinger operators have been derived on  
mathematical grounds. We refer to the textbook accounts  
\cite{CyconFKS-87, Kirsch-89a,CarmonaL-1990, PasturF-1992,  
Stollmann-2001} and the references cited therein.  
  
This paper carries over the fundamental properties of random
Schr\"odinger operators to random Laplace-Beltrami operators,
i.e.~Laplacians with random metrics. Namely, we
  
\begin{enumerate}[(A)]  
\item  
discuss a framework for random operators on manifolds with randomness
entering both via potential and metrics,
\item  
show measurability of the introduced operators, which implies, in
particular, almost sure constancy of their spectral features,
\item  
prove existence and \emph{selfaveraging} property of the
\emph{integrated density of states} together with a \v{S}ubin type
trace formula.
\end{enumerate}  
Thereby we extend and apply the earlier
\cite{PeyerimhoffV-2002,LenzPV-2002}.  The main result of this paper
is result (C) concerning the integrated density of states (IDS).
Physically, the integrated density of states measures the number of
electron energy levels per unit volume up to a given energy value. It
can be obtained by a macroscopic limit, where ergodicity of the family
of operators yields the selfaveraging nature, i.e.~the non-randomness,
of this quantity. It is sometimes called \emph{spectral density
function}.
\smallskip

Let us put these results in perspective. Probably the most prominent
success of the theory of random Schr\"odinger operators is the proof
of {\em localization}.  This phenomenon has been explained on physical
grounds by Anderson \cite{Anderson-58}, but only in the late 70ties
first rigorous results were established, see the original papers
\cite{GoldsheidMP-77,FroehlichS-83,AizenmanM-93} or
\cite{Stollmann-2001} for a monograph exposition.
  
In \cite{Davies-1990a} Davies studies among others the relation of
heat kernels on a manifold associated to different metrics. In this
context he raises the question of localization due to random
metrics. This should be analogous to the phenomena occurring in
quantum wave guides \cite{KleespiesS-2000}.

In comparison to localization the study of the integrated density of
states undertaken in this paper is a physically more basic and
technically less involved question.  Still this quantity comprises
many important spectral features of the random Schr\"odinger operator
and its understanding can be seen as a first step towards the proof of
localization.  Namely, the multiscale proof of localization of
Fr\"ohlich and Spencer \cite{FroehlichS-83} derived for (specific
models of) random Schr\"odinger operators in Euclidean space relies on
the continuity and asymptotic properties of the integrated density of
states. These were first studied by Wegner \cite{Wegner-81},
respectively by Lifshitz \cite{Lifshitz-1964}. In a forthcoming paper
\cite{LenzPPV}, we derive results on the (dis)continuity of the IDS for
periodic and random operators on manifolds. There, we furthermore
discuss some similarities and differences between random
Laplace-Beltrami operators and divergence type operators. For the time
being, let us only emphasize that Euclidean random divergence type operators 
do not cover our models, due to the more general geometry and
underlying group structure we consider.
% This is even true in the Euclidean case $(X,\Gamma) = (\RR^n,\ZZ^n)$.

\smallskip  
  
The paper is organized as follows. In the next section we introduce
our model and state the main results. In Section \ref{QFM} we
introduce quadratic forms and derive the measurability of the
quantities we are considering, thereby giving a precise form to (B)
above. Section \ref{absResults} is devoted to general results on
random operators which are proven in an abstract setting in
\cite{LenzPV-2002}. This presents our treatment of (A) above. A
discussion of heat kernels on manifolds is given in Section \ref{HK},
specializing to the principle of not feeling the boundary in Section
\ref{PNFB}. We derive uniform bounds for the kernels of the semigroups
of a random family of Schr\"odinger operators acting on a manifold and
including singular nonnegative potentials. Using these results, we then
prove our main result concerning (C) in Section \ref{construct}.

\section{Model and results} \label{ModRes}  

In this section we state the main results about the existence and
non-randomness of the integrated density of states. Beforehand we
explain the geometric setting we are working in and the properties of
the random Schr\"odinger operator.
\medskip

Consider a complete Riemannian manifold $X$ of dimension $n$ with
metric $g_0$ and associated volume form $\vol_0$. Let $\Gamma$ be a
discrete infinite subgroup of the isometries of $(X,g_0)$, acting cocompactly,
freely and properly discontinuously on $X$. Consequently, $M =
X/\Gamma$ is a compact Riemannian manifold. Furthermore, let $(\Omega,
\calB_\Omega, \PP)$ be a probability space on which $\Gamma$ acts
{\em ergodically} by measure preserving transformations.
  
\begin{definition}  
\label{randommetric}  
A family of Riemannian metrics $\{g_\omega\}_{\omega\in\Omega}$ on $X$
with corresponding volume forms $\vol_\omega$ is called a {\em random
metric on $(X,g_0)$} if the following properties are satisfied:
\begin{enumerate}[(M1)]  
\item  
The map $\Omega \times TM \to \RR$, $(\omega,v) \mapsto g_\omega(v,v)$ 
is jointly measurable.  
\item  
There is a $C_g \in \, ]0,\infty[$ such that  
\begin{equation*} \label{quasiisom}  
C_g^{-1} g_0(v,v) \le g_\omega(v,v) \le C_g  g_0(v,v) \  \mbox{for all} \,  
\, v \in TX.  
\end{equation*}  
\item  
There is a $C_\rho > 0$ such that  
\[ \vert \nabla_0 \rho_\omega(x) \vert_0 \le C_\rho \  \mbox{for all} \,
\, x \in X, \] 
where $\nabla_0$ denotes the gradient w.r.t $g_0$,
$\rho_\omega$ is the unique smooth density satisfying $d\vol_0 = \rho_\omega
d\vol_\omega$, and $\vert v \vert_0^2 = g_0(v,v)$.
\item  
There is a uniform lower bound $K \in \RR$ for the Ricci curvatures of all
Riemannian manifolds $(X,g_\omega)$.   
\item
The metrics are compatible in the sense that the deck transformations  
\[  
\gamma\colon (X,g_\omega) \to (X,g_{\gamma \omega}) ,\quad \gamma
\colon x \mapsto \gamma x
\]  
are isometries.
\end{enumerate}
\end{definition} 
\hspace*{-.5em}(M5) implies that, in particular, the induced maps
$U_{(\omega,\gamma)}\colon L^2(X,\vol_{\gamma^{-1}\omega}) \to
L^2(X,\vol_\omega)$, $(U_{(\omega,\gamma)} f)(x) = f(\gamma^{-1}x)$
are unitary operators.
  
Based on this geometric setting, we consider a family of Schr\"odinger
operators. These operators are defined via quadratic forms, as
explained in Section \ref{QFM}.
  
\begin{definition}  
\label{randomoperator}  
Let $\{ g_\omega \}$ be a random metric on $(X,g_0)$. For each
$\omega\in\Omega$ let $H_\omega = \Delta_\omega + V_\omega$ be a
Schr\"odinger operator defined on the Hilbert space
$L^2(X,\vol_\omega)$. $\{H_\omega\}_{\omega\in\Omega}$ is called a
{\em random (Schr\"odinger) operator} if it satisfies the following {\em
equivariance} condition
\begin{equation} \label{compcomp}  
H_\omega = U_{(\omega,\gamma)} H_{\gamma^{-1} \omega} U_{(\omega,\gamma)}^*,  
\end{equation}  
for all $\gamma \in \Gamma$ and $\omega \in \Omega$, and if the
potential $V\colon \Omega \times X\to \RR$ is jointly measurable,
nonnegative and $V_\omega = V(\omega,\cdot) \in L_{loc}^1(X)$, for all
$\omega \in \Omega$.
\end{definition}  

For technical reasons we require that the $\sigma$-algebra $\calB_\Omega$
is countably generated. This can always be established by changing to
an equivalent version of the defining stochastic processes given by
the random potential and the random metric. This has been done for the
potential explicitely in Remark 2.8 of \cite{LenzPV-2002}.

In Section \ref{QFM} we extend the standard notion of measurability
for a family of operators acting on a fixed Hilbert space
\cite{KirschM-82a} to operators acting on varying Hilbert spaces. This
leads to the fundamental
    
\begin{theorem}\label{measur}  
A random operator $\{H_\omega\}_{\omega\in\Omega}$ is a measurable
family of operators.
\end{theorem}  
  
From this theorem and the results of \cite{LenzPV-2002} we immediately
obtain the following result. (Note that $\sigma_{pp} $ denotes the
{\em closure} of the set of eigenvalues.)

\begin{theorem} \label{constancy}  
There exist $\Omega' \subset \Omega$ of full measure and $\Sigma,
 \Sigma_\bullet\subset \RR$, such that $ \sigma(H_\omega)=\Sigma,
 \quad \sigma_\bullet (H_\omega)= \Sigma_\bullet\quad \text{ for all
 }\omega\in \Omega' $ where $\bullet = disc,ess,ac, sc,pp$.  Moreover,
 $\Sigma_{disc}=\emptyset$.
\end{theorem}  
  
The above two theorems and the framework underlying their proofs
complete our investigation of (A) and (B) of the introduction.
  
Next, we introduce the {\em (abstract) density of states} for a
random operator $\{ H_\omega \}$ as the measure on $\RR$, given by
\begin{equation} \label{absIDS}
\rho_H(f):=  
\frac{\EE \l[ \tr \left( \chi_\calF f(H_\bullet) \right) \r]  
}{\EE \l[ \vol_\bullet(\calF)\r] }, \quad \text{$f$ bounded, measurable.} 
\end{equation}
Here $\calF \subset X$ is a precompact $\Gamma$-fundamental domain
with piecewise smooth boundary, $\EE$ denotes the expectation with
respect to $\PP$ and $\tr=\tr_\omega$ is the trace on the Hilbert space 
$L^2(X,\vol_\omega)$, where we suppress the index $\omega$ in the following.
The expression \eqref{absIDS} is closely related to a trace $\tau$ of a von Neumann algebra, as discussed in Section
\ref{absResults} and summarized in

\begin{theorem}\label{AbstIDS}  
$\rho_H$ is a spectral measure for the direct integral operator 
\[ H:= {\int_\Omega}^\oplus H_\omega \, d \PP(\omega)\]
and $\rho_H (f)$ and $\tau(f(H))$ coincide for arbitrary bounded
measurable $f$ on $\RR$, up to a fixed constant factor. In particular,
the almost sure spectrum $\Sigma$ coincides with the topological support
$\{ \lambda\in \RR : \rho(\, ]\lambda-\epsilon, \lambda + \epsilon[\, )>0
\quad \text{for all $\epsilon >0$}\}$
of $\rho_H$.
\end{theorem}
Recall that a measure $\phi$ on $\RR$ is a spectral measure
for selfadjoint operator $H$ with spectral family $E_H$ if, for Borel
measurable $B\subset\RR$, $\phi(B)=0 \Leftrightarrow E_H (B)=0$.
\smallskip

To state our main result (see (C) of the introduction), it is
indispensable to assume that the underlying discrete group $\Gamma$ is
{\em amenable}. We introduce restrictions of operators on $X$ to open
sets $D\subset X$ with $\vol_\omega (D) <\infty$. The restriction of
$H_\omega$ to $D$ with Dirichlet boundary conditions (b.c.) will be
denoted by $H_\omega^D$. The restriction $H_\omega^D$ is again
selfadjoint, bounded below and has purely discrete
spectrum. Therefore, we may enumerate its eigenvalues in increasing
order, counting multiplicities: $\lambda_1 (H_\omega^D) \le \lambda_2
(H_\omega^D) \le \dots \lambda_i (H_\omega^D) \to \infty$. We define
the {\em normalized eigenvalue counting function} as
\begin{equation} \label{ewcfunc}  
N_\omega^D(\lambda) = \frac{\# \{ i \mid \lambda_i(H_\omega^{D}) < \lambda  
 \}}{\vol_\omega(D) }. \  
\end{equation}  
$N_\omega^D$ is a distribution function and has countably many
discontinuity points.
  
Amenability of $\Gamma$ guarantees the existence of an exhaustion of
$X$ by open sets $\{D^j\}_j$ with very strong additional properties,
see Section \ref{construct}. Such an exhaustion $\{D^j\}_j$ is called
an {\em admissible sequence} of subsets of $X$. For the associated
restricted operators we use the shorthand $H_\omega^j=H_\omega^{D^j}$
and, similarly, $N_\omega^j= N_\omega^{D^j}$. Our main result
establishes the selfaveraging property of the IDS and expresses it by
a \v Subin type trace formula \cite{Shubin-1979,Shubin-1982}:
\begin{theorem}\label{selfaverIDS}  
Let $\{D^j\}_j$ be an admissible sequence and $\{H_\omega\}_\omega$ be as
above. There exists a set $\Omega'$ of full measure such that
\[  
\lim_{j\to\infty} N_\omega^{j}(\lambda) = \rho_H(\, ]-\infty,\lambda[\, ),  
\]  
for every $\omega \in \Omega'$ and every point $\lambda\in \RR$ with
$\rho_H(\{ \lambda\})=0$.
\end{theorem}  
The distribution function of $\rho_H$ is denoted by $N_H$ and is
called the {\em integrated density of states (IDS)} of the random
operator $\{ H_\omega \}$, i.e., 
\[ 
N_H(\lambda) = \rho_H(\, ]-\infty,\lambda[\, ). 
\]
Since $N_H$ can be obtained by an exhaustion procedure $D^j \to X$
without integrating over $\Omega$ explicitly, it is called {\em
selfaveraging}. The proof of Theorem \ref{selfaverIDS} in Section
\ref{construct} actually also establishes the following result. For a
set $D^j \subset X$ in an admissible sequence denote
\begin{equation} \label{NECFnoBC}  
N_\omega^{j,f}(\lambda):=  
\frac{ \tr \left( \chi_{D^j} E_{\omega}(\lambda) \right)}{\vol_\omega(D^j) }.  
\end{equation} 
Here, the superscript $f$ stands for the fact that this finite volume
IDS is defined without the use (i.e.~{\em free}) of boundary conditions.

\begin{coro}  
For almost every $\omega \in \Omega$ the convergence
$\lim_{j\to\infty} N_\omega^{j,f}(\lambda) = N_H (\lambda)$ holds at
every continuity point $\lambda$ of $N_H$.
\end{coro}  

This means that in the macroscopic limit $D^j\to X$ it is not felt
whether the restriction of the operator in space, or
the projection on an energy  interval took place first.

For simplicity, we have so far assumed the potentials $V$ to be
nonnegative. It suffices to assume that the $V_\omega$ are uniformly
bounded below by a constant $C$ not depending on $\omega\in \Omega$.
Then our results apply to the shifted operator family $\{H_\omega
-C\}_{\omega\in\Omega}$. Since
$ N_{H-C}(\lambda-C) = N_H(\lambda)$ for all $ \lambda \in \RR$, 
and similarly for the normalized eigenvalue counting functions,  
the results carry over to the original operators.

\section{Quadratic forms and measurability\label{QFM}}  

In this section we give a precise definition of the operators we are
dealing with and show their measurability.
\medskip  
  
To introduce our operators, we will use quadratic forms. The relevant
theory can be found, e.g., in the first two sections of \cite{Davies-1989}. It is
developed there for $X=\RR^n$ but carries directly over to arbitrary
manifolds $X$.
  
We abbreviate the scalar product in the tangent space by
\[  
\la v,w\ra_\omega := g_\omega(x)(v,w) \text{ for all } v,w \in
T_xX.
\]  
For $D \subset X$ open and each $\omega\in \Omega$ we define the
quadratic forms
\[  
 \widetilde{Q} (\Delta_\omega^D) \colon \CcsmoothD \times \CcsmoothD
\rightarrow \RR, \; \: (f,h)\mapsto \int_D \la\nabla f (x), \nabla
h(x)\ra_\omega d\vol_\omega (x)
\]  
and  
\[  
\widetilde{Q} (V_\omega^D) \colon \CcsmoothD \times \CcsmoothD
\rightarrow \RR, \; \: (f,h)\mapsto \int_D f(x) V_\omega(x) h(x) d\vol_\omega
(x).
\]  
These forms are closable and their closures $Q(\Delta_\omega^D)$ and
$Q(V_\omega^D)$, respectively, give rise to selfadjoint nonnegative
operators $\Delta_\omega^D$ and $V_\omega$. Next, consider the form
\[
\widetilde{Q} (\HN^D)\colon \CcsmoothD \times \CcsmoothD \rightarrow
\RR, \; \: (f,h)\mapsto Q(\Delta_\omega^D)(f,h) + Q(V_\omega^D)(f,h).
\]
This form is closable, the closure $Q(\HN^D)$ is the form sum of
$Q(\Delta_\omega^D)$ and $Q(V_\omega^D)$, and $Q(\HN^D)$ induces,
again, a selfadjoint operator (see \cite[Thm.~1.8.1]{Davies-1989}). Of
course, for smooth $V$ and $f\in \CcsmoothD$ we have $\HN^D f (x) =
\Delta_\omega f (x) + V(x) f(x)$. The form $Q(\HN^D)$ is a Dirichlet
form (see \cite[Thm.~1.3.5]{Davies-1989}), i.e., it satisfies
\[
\ExpH \colon L^{\infty} (D,\vol_\omega)\rightarrow
L^{\infty}(D,\vol_\omega) \;\: \mbox{is a contraction for every $t>0$}
\]
and  
\[
\ExpH \colon L^{2} (D,\vol_\omega)\rightarrow
L^{2}(D,\vol_\omega) \mbox{ is positivity preserving for every $t>0$}.
\]
Semigroups $e^{-t H}$ associated to Dirichlet forms are called
symmetric Markov semigroups.  

\medskip  

There exist positive, smooth functions $\rho_\omega \in C^\infty(X)$
such that
\[  
\int_X f(x) d\vol_0(x) = \int_X f(x) \rho_\omega(x) d\vol_\omega(x).
\]  
More explicitly, $\rho_\omega(x)$ is given by
\[  
\rho_\omega(x) = \l(\det g_0(e_i^\omega, e_j^\omega) \r)^{1/2} =  
\l(\det g_\omega(e_i^0, e_j^0)\r)^{-1/2},  
\]  
where $e_1^0,\dots,e_d^0 \in T_xX$ is any base of $T_xX$ orthonormal
w.r.t.  $g_0$ and $e_1^\omega,\dots,e_d^\omega \in T_xX$ is any base
orthonormal w.r.t. $g_\omega$. Consequently, the operators
\[ S_\omega\colon L^2(D,\vol_0) \to L^2(D,\vol_\omega), \quad  
S_\omega(f) = \rho_\omega^{1/2} f \] are unitary. The $L^2$-products
on $L^2(D,\vol_0)$ and on $L^2(D,\vol_\omega)$ are denoted by
$(\cdot,\cdot)_0$ and $(\cdot, \cdot)_\omega$, respectively. The
corresponding norms are denoted by $\Vert \cdot \Vert_0$ and $\Vert
\cdot \Vert_\omega$.  

It follows from property (M2) of Definition \ref{randommetric} that  
\begin {equation}  
\label{rhoest}  
C_g^{-n/2} \le \rho_\omega(x) \le C_g^{n/2} \ \ \mbox{for all} \, x \in D, \ 
 \omega\in \Omega. 
\end{equation}  
%which in turn, together with property (M3) and the chain rule, implies  
%\[  
%\vert \nabla_0 \rho_\omega^{\, \pm \, 1/2}(x) \vert_0 \le C_g^{3n/4}
%\, \vert \nabla_0 \rho_\omega(x) \vert_0 \ \ \mbox{for all} \, x \in
%X, \ \omega\in \Omega.
%\]  
Now we introduce the notion of {\em measurability} of a family of
selfadjoint operators, indexed by the elements of $\Omega$. It is a
modification of the definition from \cite{KirschM-82a} to operators
with varying domains of definition.

\begin{definition}  
\label{measurableH}  
A family of selfadjoint operators $\{H_\omega\}_\omega$, where the
domain of $H_\omega$ is a dense subspace $\calD_\omega$ of $L^2(D, \vol_\omega)$, is
called a {\em measurable family of operators} if
\begin{equation}  \label{weakmeas}  
\omega\mapsto ( f_\omega, F(H_\omega) f_\omega)_\omega
\end{equation}
is measurable for all bounded, measurable functions $F\colon \RR\rightarrow \CC$  
and all $f \colon \Omega\times D\rightarrow \RR$ measurable with $f_\omega \in L^2(D,\vol_\omega)$, $f_\omega(x) = f(\omega,x)$, for every $\omega\in \Omega$.
\end{definition}  

\begin{remark}
In our setting, due to (M2), the above definition can be slightly
simplified. Namely, a family of operators $\{H_\omega\}_\omega$ is
measurable if and only if
\begin{equation}  
\label{wenigermeas} 
\omega\mapsto ( f , F(H_\omega) f )_\omega \text{ is measurable }
\end{equation} 
for all $F\colon \RR\rightarrow \CC, \, F \in L^\infty$ and all $f \in
L^2(D, \vol_0)$. (Note that, due to (M2), $L^2(D, \vol_0)$ and $L^2(D,
\vol_\omega)$ coincide for all $\omega \in \Omega$ as sets, though not
in their scalar product.)

To see this, note that \eqref{wenigermeas} implies the same statement
for $f(\omega,x)$ replaced by $h(\omega,x)= g(\omega) f(x)$ where $g
\in L^2(\Omega)$ and $f \in L^2(D, \vol_0)$. Such functions form a
total set in $L^2(\Omega\times D, \PP\circ \vol)$.

Now, consider a measurable $h \colon \Omega\times D\rightarrow \RR$
such that $h_\omega:= h(\omega, \cdot) \in L^2(D,\vol_\omega)$ for
every $\omega\in \Omega$.  Then $h^n(\omega,x):= \chi_{h,n}(\omega)
h(\omega,x)$ is in $L^2(\Omega\times D, \PP\circ \vol)$ where
$\chi_{h,n}$ denotes the characteristic function of the set $\{\omega|
\, \|h_\omega\|_{L^2(D,\vol_\omega)} \le n \} $. Since $\chi_{h,n} \to
1$ pointwise on $\Omega$ for $n \to \infty$ we obtain
\[
( h^n_\omega, F(H_\omega) h^n_\omega)_\omega \to ( h_\omega,
F(H_\omega) h_\omega)_\omega
\]
which shows that $\{H_\omega\}_\omega$ is a measurable family of operators.
\end{remark}  

The following proposition (and its proof) is a variant of Proposition 3
in \cite{KirschM-82a}. It suits our purposes and shows that our notion
of measurability is compatible with theirs: Let $\{A_\omega\}_\omega$
be a family of densely defined nonnegative selfadjoint operators on a
fixed Hilbert space $\calH$.  Denote by $\tilde\Sigma=
\overline{\bigcup_\omega \sigma(A_\omega)}$ the closure of all spectra
and by $\calF_i$ the the following classes of functions: $\calF_1= \{
\chi_{]-\infty,\lambda[} | \, \lambda \ge 0 \}$, $\calF_2= \{ x
\mapsto e^{itx} | \, t \in \RR \}$, $\calF_3= \{ x \mapsto e^{-tx} |
\, t \ge 0 \}$, $\calF_4= \{ x \mapsto (z-x)^{-1} | \, z\in
\CC\setminus\tilde\Sigma \}$, $\calF_5=\calF_4(z_0)= \{ x \mapsto
(z_0-x)^{-1} \}$ for a fixed $z_0\in \CC\setminus\tilde\Sigma $,
$\calF_6= C_b =\{ f\colon \RR \to \CC | \, f \text{ bounded,
continuous} \}$, and $\calF_7= L^\infty =\{ f\colon \RR \to \CC | \, f
\text{ bounded, measurable} \}$.

\begin{prop}  
\label{equivProp}  
%   \marginpar{Achtung: um LPV-I anwenden zu koennen brauchen wir die 
%   Formel \eqref{weakmeas}  mit $f$ bei denen die Faser mit $\omega$ 
%   variieren kann. Der Rest dieses Abschnitts beweist \eqref{weakmeas}  
%   nur fuer $f_\omega\equiv f$.} 
The following properties are equivalent:  
\begin{equation*} 
({{\mathbf F}_i}) \hspace{4em} \omega \mapsto \la f, F(A_\omega)h
\ra_{\, \calH} \text{ is measurable for all } f,h \in \calH \text{ and
} F \in \calF_i,
\end{equation*} 
where $ i =1,\dots,7$.  
\end{prop}

\begin{proof}  

For the equivalence of (\textbf{F}$_4$) and (\textbf{F}$_5$) we assume
$d(z_0,\tilde\Sigma) =\delta$ and that $(z_0-H_\omega)^{-1}$ is weakly
measurable. Using a Neumann series expansion as in \cite[Theorem
VI.5]{ReedS-80} one infers the weak measurability of
$(z-H_\omega)^{-1}$ for all $z$ with $d(z, z_0)<\delta$. Iterating
this argument, we obtain measurability of $(z-H_\omega)^{-1}$ for all
$z \in \CC \backslash \tilde \Sigma$.

Now, by the Stone/Weierstrass theorem we obtain the equivalence
of (\textbf{F}$_2$), (\textbf{F}$_3$), (\textbf{F}$_4$), (\textbf{F}$_5$),
(\textbf{F}$_6$).

The equivalence of (\textbf{F}$_1$) and (\textbf{F}$_7$) follows by
monotone class arguments. 

As (\textbf{F}$_7$) $\Rightarrow$ (\textbf{F}$_6$) is clear, it only
remains to prove (\textbf{F}$_6$) $\Rightarrow$ (\textbf{F}$_1$). This
is immediate as every characteristic function $\chi_{]-\infty,\lambda[}$ is
a pointwise monotone limit of continuous functions. 
\end{proof}

We prove now that the random operator $\{ H_\omega \}$ introduced in
Section \ref{ModRes} is measurable in the sense of Definition
\ref{measurableH}. The first step in the proof is to pull all
operators $\{H_\omega\}_\omega$ on the same Hilbert space by the
unitary transformation $S_\omega$ and to show the following
comparability property of the associated quadratic forms:

\begin{prop} \label{glmBProp}  
Let the selfadjoint operators 
\[ A_\omega \colon (S_\omega)^{-1} \calD(\Delta_\omega^D) \subset L^2 (D,\vol_0) 
\longrightarrow L^2(D,\vol_0) \] 
be defined by $A_\omega := (S_\omega)^{-1} \Delta_\omega^D S_\omega$.
Let $Q_0, Q_\omega$ be the quadratic forms associated to the operators
$\Delta_0^D$ and $A_\omega$. Then there is a constant $C_A$ such that
\begin {equation} \label{QomQ0}  
C_A^{-1} \left( Q_0(f,f) + \Vert f \Vert_0^2 \right)\le  
Q_\omega(f,f) + \Vert f \Vert_0^2 \le  
 C_A \left( Q_0(f,f) + \Vert f \Vert_0^2 \right).  
\end{equation}  
for all $f \in C_c^\infty(D)$ and $\omega\in\Omega$. Moreover, there
exists a dense subspace $\calD \subset L^2(D,\vol_0)$ with $\calD=
\calD (A_\omega^{\frac{1}{2}}) = \calD ((\Delta_0^D)^{\frac{1}{2}}) $
for every $\omega \in \Omega$ and \eqref{QomQ0} holds for every $f\in
\calD$.
\end{prop}  

\begin{proof}  
Direct calculation for $f \in C_c^\infty(D)$ shows  
\begin {equation}  
Q_\omega(f,f) = (S_\omega f, \Delta_\omega S_\omega f)_\omega \le 2
\left( \Vert \rho_\omega^{1/2} \nabla_\omega f \Vert_\omega^2 + \Vert f
\nabla_\omega \rho_\omega^{1/2} \Vert_\omega^2 \right).
\label{Qomest}  
\end{equation}  
To bound $\Vert \rho_\omega^{1/2} \nabla_\omega f \Vert_\omega^2$ we
consider the $n \times n$-matrix $A = (a_{ij})$ defined by $ e_i^0 =
\sum_{j=1}^n a_{ij} \,e_j^\omega$. A calculation using (M2) in
Definition \ref{randommetric} shows $C_g^{-1} \le A A^\top \le
C_g$. This implies
\begin{equation} 
\label{gradest}  
C_g^{-1} \vert \nabla_\omega f(x) \vert_\omega^2 \le \vert \nabla_0
 f(x) \vert_0^2 \le C_g \vert \nabla_\omega f(x) \vert_\omega^2.
\end{equation} 
and thus  
\[  
\Vert \rho_\omega^{1/2} \nabla_\omega f \Vert^2_\omega = \int_D \vert
\nabla_\omega f(x) \vert_\omega^2 d\vol_0(x) \le C_g \int_D \vert \nabla_0
f(x) \vert_0^2 d\vol_0(x) = C_g \, Q_0(f,f).
\]
To estimate $\Vert f \nabla_\omega \rho_\omega^{1/2} \Vert_\omega^2$
we use \eqref{rhoest}, \eqref{gradest} and (M3) of Definition
\ref{randommetric} to calculate
\begin{align*}  
\Vert f \nabla_\omega \rho_\omega^{1/2} \Vert_\omega^2 \le C_g^{1+n/2}
\int_D \vert \nabla_0 \rho_\omega(x) \vert_0^2 f^2(x) d\vol_\omega(x)
\le C_g^{1+n} C_\rho^2 \Vert f \Vert^2_0.
\end{align*} 
By symmetry, there is also an estimate of the form  
\[  
C_A^{-1} \left( Q_0(f,f) + \Vert f \Vert_0^2 \right) \le  
Q_\omega(f,f) + \Vert f \Vert_\omega^2,  
\]  
for all $f \in C^\infty_c(D)$, and the first statement is proven.
The statement follows now from \eqref{QomQ0}, as $C^\infty_c(D)$ is a core for
$Q_0$ and $Q_\omega$.
\end{proof}  
  
%The proposition enables us to use the following result.
    
\begin{prop}[see Prop.~1.2.6.~in \cite{Stollmann-2001}] \label{stoll}  
Let $Q_\omega$, $\omega \in \Omega$ and $Q_0$ be nonnegative {\em
closed} quadratic forms with the following properties:
\begin{itemize}  
\item[(P1)] $Q_\omega$, $\omega \in \Omega$ and $Q_0$ are defined on
the same dense subset $\calD$ of a fixed Hilbert space $\calH$.
\item[(P2)] There is a fixed constant $C > 0$ such that
\[  
C^{-1} \left( Q_0(f,f) + \Vert f \Vert_0^2 \right) \le  
Q_\omega(f,f) + \Vert f \Vert_0^2 \le C \left( Q_0(f,f) + \Vert f 
\Vert_0^2 \right).  
\]  
\item[(P3)] The map $\omega \mapsto Q_\omega(f,f)$ is measurable, for
every $f \in \calD$.
\end{itemize}  
Then the family $\{H_\omega\}_\omega$ of associated selfadjoint
operators satisfies the equivalent properties of Proposition
\ref{equivProp}.
\end{prop}

The foregoing propositions allow us to show the following:
  
\begin{prop} 
\label{Ameas}
The family \label{measurprop} $\{ A_\omega \}_\omega$ of 
Proposition \ref{glmBProp} is a measurable family of operators.
\end{prop}

\begin{proof}%[Proof of Proposition \ref{Ameas}] 
 Since $C^\infty_c (D)$ is a core for $Q_\omega$ for all $\omega$,
 the closures of this set with respect to one of the equivalent norms in \eqref{QomQ0} 
 coincide, which shows assumption (P1) of Proposition \ref{stoll}. (P2) is just \eqref{QomQ0} 
 and  (P3) is obvious for $f\in C^\infty_c (D)$. It then
 follows by approximation for all $f\in \calD$ 
\end{proof}

\begin{proof}[Proof of Theorem \ref{measur}] 
 For $n\in \NN$ and $\omega \in \Omega$, define bounded functions
 $V_\omega^n \colon X\rightarrow \RR$ by $V_\omega^n (x) :=
 \min\{n, V_\omega (x)\}$.  
 Thus, the operator sum $A_\omega^n := A_\omega + V_\omega^n$ is well
 defined, where $A_\omega$ is as in Proposition \ref{glmBProp} and $D=X$.
 Moreover, by \cite[Prop. 2.4]{KirschM-82a} and Proposition
 \ref{measurprop}, the family of operators $A_\omega^n$ is
 measurable. In particular, the corresponding semigroups
 $\omega\mapsto \exp(- t A_\omega^n)$, $t>0$, are weakly measurable.
 Now, obviously, the forms of $A_\omega^n$ converge monotonously
 towards the form of $A_\omega^\infty:= A_\omega +V_\omega$. By
 \cite[Thms. VIII.3.13a and IX.2.16]{Kato-80}, this implies that the
 semigroups of $A_\omega^n$ converge weakly towards the semigroup
 $\omega \mapsto \exp(-t A_\omega^\infty)$ for $n\to \infty$, and the
 measurability of the family $A_\omega^\infty$ follows. Finally, this
 implies measurability of the family $H_\omega$, since $H_\omega =
 S_\omega A_\omega^{\infty} {S_\omega}^{-1}$ and $S_\omega$ is
 multiplication with the measurable function $(x,\omega) \mapsto
 \rho_\omega (x)$.
\end{proof}  
  
The same arguments show measurability
of the restricted operators $\{ H_\omega^D \}_\omega$.

\section{Abstract spectral properties of random operators\label{absResults}}  

We saw in the last section that a random operator $\{H_\omega\}_\omega$ is a
measurable family of operators. This enables us to make use of the
results derived in \cite{LenzPV-2002} for random operators in an
abstract setting. The following information can be inferred from the
cited source.
  
\begin{definition}  
A family $\{A_\omega\}_{\omega\in \Omega}$ of bounded operators
$A_\omega\colon L^2(X,\vol_\omega)\to L^2(X,\vol_\omega)$ is called a
{\em bounded random operator} if it satisfies:
\begin{enumerate}[\rm (i)]  
\item $\omega\mapsto \langle g_\omega, A_\omega f_\omega\rangle$ is
measurable for arbitrary $f,g\in L^2(\Omega\times X, \PP\circ \vol)$.
\item There exists a $C\geq 0$ with $\|A_\omega\|\leq C$ for almost
all $\omega \in \Omega$.
\item For all $\omega\in\Omega, \gamma \in \Gamma$ the equivariance
condition $ A_\omega = U_{(\omega,\gamma)} A_{\gamma^{-1} \omega}
U_{(\omega,\gamma)}^*$is satisfied.
\end{enumerate} 
\end{definition}
 
Two bounded random operators $\{A_\omega\}_\omega,
\{B_\omega\}_\omega$ are called {\em equivalent}, $\{A_\omega\}_\omega\sim
\{B_\omega\}_\omega$, if $A_\omega=B_\omega$ for $\PP$-almost every
$\omega\in \Omega$.  Each equivalence class of bounded random
operators $\{A_\omega\}_\omega$ gives rise to a bounded operator $A$
on $L^2(\Omega\times X, \PP\circ \vol)$ by $(A f) (\omega,x) :=
A_{\omega} f_{\omega}(x)$, see Appendix A in \cite{LenzPV-2002}.  This
allows us to identify the equivalence class of $\{A_\omega\}_\omega$ with the
bounded operator $A$.  

By \eqref{compcomp} and the last section, the resolvents, spectral
projections and the semigroup associated to $\{H_\omega\}_\omega$ are
all bounded random operators. Theorem 3.1 in \cite{LenzPV-2002} states
that the set of bounded random operators forms a von Neumann algebra
$\calN$.  Choose a measurable $ u\colon\Omega\times X \to \RR^+$ with
$ \sum_{\gamma \in\Gamma} u_{\gamma^{-1}\omega}(\gamma^{-1}x)\equiv 1
$ on $\Omega\times X$ and define the mapping
\[  
\tau(A) :=   \EE \l[ \tr (u_\bullet A_\bullet) \r]  
\]  
on the set of non-negative operators in $\calN$. This $\tau$ is
independent of $u$ (chosen as above) and defines a trace on $\calN$ of
type II$_\infty$, which is closely related to the IDS.
Namely, the spectral projections $\{E_\omega(\lambda)\}_\omega$ onto
the interval $]-\infty, \lambda[$ of a random operator
$\{H_\omega\}_\omega$ form a bounded random operator. Thus it is an
element of $\calN$ and agrees with the spectral projection of $H:=
\int_\Omega^\oplus H_\omega \, d \PP(\omega) $ onto $]-\infty,
\lambda[$.  Hence $\tau(E(\lambda))$ is well defined and the
choice $u_\omega(x) = \chi_{\calF} (x)$ yields the identity $\tau(E(\lambda))=
\EE \, (\vol_\bullet \calF) N_H(\lambda)$, where $\calF$ is a fundamental
domain as discussed after \eqref{absIDS}.

\smallskip  
  
Now, Theorems \ref{constancy} and \ref{AbstIDS} follow from Sections 4
and 5 of \cite{LenzPV-2002}.

\section{Heat kernels}\label{HK}  
  
In this section we investigate existence and properties of the kernels of
the semigroups $\ExpH$ and $\ExpHD$. It will be of particular
importance to us to keep track of the dependence of the estimates both
on the potential and the metric, since they vary with the random
parameter $\omega\in \Omega$.
\medskip  
  
We start with the kernels of the Laplacians $\Delta_\omega$. Sobolev
embedding theorems and spectral calculus directly show that
$$ \exp(-t \Delta_\omega) \colon L^2 (X,\vol_\omega) \longrightarrow
L^{\infty} (X,\vol_\omega) \;\: \mbox{is bounded for every $t>0$}.$$
Thus, $\exp(-t \Delta_\omega)$ is ultracontractive, and, by \cite[Lemma
2.1.2]{Davies-1989}, this implies that $\exp(-t \Delta_\omega)$ has a
kernel $k_{\Delta_\omega}$ with
\begin{equation} \label{ctfirst}
0\leq k_{\Delta_\omega} (t,x,y) \leq \Vert \exp(-t \Delta_\omega)
\Vert_{1,\infty} =: C_t^\omega, \quad \text{for almost all $x,y \in X$,}
\end{equation}
where $\|A\|_{1,\infty}$ denotes the norm of $A \colon
L^1\longrightarrow L^\infty$. By the Trotter product formula we see
that, for $f\geq 0$, $f\in L^1(X,\vol_\omega)$,
$$
0\leq \ExpH f(x) \leq \exp(-t\Delta_\omega) f(x) \leq C_t^\omega \|f\|_{L^1}
$$
for almost every $x\in X$. Thus, $\ExpH \colon L^1 (X,\vol_\omega)
\rightarrow L^\infty (X,\vol_\omega) $ is also bounded by $C_t^\omega$ and
we have
\begin{equation} \label{bounds}  
0\leq \ko (t,x,y) \leq C_t^\omega \;\:\mbox{for almost every $x,y\in
X$}.
\end{equation}  
To obtain a better estimate, we show that the $L^2$-kernel of the heat
semigroup coincides with the fundamental solution of the heat equation 
%$\frac{d}{dt} u + \Delta +Wu = 0$, 
as defined, e.g., in \cite{Dodziuk-1983} or \cite{Chavel-84} in the case of the pure 
Laplacian. 
This allows us to apply estimates
of \cite{LiY-1986} for the fundamental solution. Uniqueness of the
fundamental solution and its agreement with the $L^2$-kernel are
well-known (see, e.g., \cite{Dodziuk-1983}). For completeness reasons,
we give a short alternative functional analytic proof of this
agreement based on a theorem of \cite{Davies-1989} in the more general
case with a smooth potential $W$.

\begin{theorem}\label{allesgleich} Let $\{H_\omega\}_\omega = \Delta_\omega 
+ W_\omega$ be a random operator with smooth potential $W_\omega \in
C^\infty(X)$. Then, the kernel $\ko$ has a {\em nonnegative} representative in
$C^{\infty} (\, ]0,\infty[\times X\times X)$. Moreover, we have:
\begin{itemize}  
\item[(HE)] $\ko$ is a solution of the heat equation: $(\frac{d}{dt} +
\Delta^y_\omega + W_\omega) \ko(t,x,y) = 0$, where
$\Delta^y_\omega$ denotes the $\Delta_\omega$ operator acting on the
variable $y$.
\item[(W)] $\ko(t,x,\cdot)$ converges weakly to the point mass in $x$:
$\int \ko (t,x,y) f(y) dy \rightarrow f(x)$, as $t\to 0$, for every
bounded continuous $f$ on $X$ and every $x\in X$.
\end{itemize}  
\end{theorem}  

\begin{proof} Mimicking the proof of \cite[Thm.~5.2.1]{Davies-1989}, we 
infer that $\ko$ has a representative in $C^{\infty} (\, ]0,\infty[\times
X\times X)$. Since $\exp(-tH_\omega)$ is positivity preserving, we
conclude that $\ko \ge 0$. Now, direct calculations show (HE). To show
(W), we recall that $\exp(-t H_\omega)$ is a contraction. Thus, by
standard measure theory (see, e.g., \cite[Satz 30.8]{Bauer-92}) it
suffices to consider only $f \in C_c(X)$. We can even further restrict
the set of functions to $f \in \CcsmoothX$, since $\CcsmoothX$ is
dense in $C_c(X)$ with respect to the $\sup$-norm. By elliptic
regularity and Sobolev embeddings, there exist $a, b >0$ and $ j \in
\NN$ with \bea \|\ExpH f - f\|_\infty \leq a \| \HN^j (\ExpH f -
f)\|_2 + b \| \ExpH f - f\|_2 \eea for every $t\geq 0$. By spectral
calculus, the right hand side tends to zero as $t \to 0$, and the
theorem is proven.
\end{proof}  

To formulate the results of Li and Yau \cite{LiY-1986} which we will be using 
we denote by $d_\omega\colon X \times X \to [0,\infty[$ the Riemannian distance function on $X$
with respect to $g_\omega$, and similarly by $d_0$ the one with respect to the metric $g_0$.

\begin{prop}\label{LiYau}  
For every $t>0$ there exist constants $C_t>0$, $\alpha_t>0$ with
\begin{equation} \label{kdelest}
k_{\Delta_\omega}(t,x,y) \leq C_t \exp \big(-\alpha_t \,d^2_\omega(x,y)\big)
\end{equation}
for all $\omega \in \Omega$. In particular, the following holds:
\begin{enumerate}[\rm (i)]  
\item  
$C_t^\omega \leq C_t$ for every $\omega\in \Omega$, where $C_t^\omega$
was defined in \eqref{ctfirst}.
\item  
For all $a > 0$, there exists a $B_{t,a}<\infty $ such that the estimate
\[
\int_X k_{\Delta_\omega}^{a}(t,x,y) d\vol_\omega (y) \leq B_{t,a}
\]
holds uniformly in $x\in X$ and $\omega\in \Omega$. We set $B_t :=
B_{t,1}$.
\end{enumerate}  
\end{prop}  

\begin{proof} Using \cite[Cor. 3.1]{LiY-1986}, property (M2) of Definition 
\ref{randommetric} and \eqref{rhoest}, we obtain the estimate
\eqref{kdelest} for the fundamental solution of the heat equation
without potential. Note that the uniform lower bound $K >-\infty $ for the
Ricci curvatures of $(X,g_\omega)$ enters into the constant $C_t$.
By Theorem \ref{allesgleich} the fundamental solution agrees with
the $L^2$-kernel of the semigroup $\exp(-t \Delta_\omega)$. Given
this estimate, (i) and (ii) are easy consequences. Note for (ii) that
the volume of metric balls of radius $r$ can be estimated (uniformly
in $\omega$) from above by $C_1 \exp(C_2 r)$ with fixed constants
$C_1, C_2 > 0$, by property (M2), \eqref{rhoest}, and the Bishop
volume comparison theorem (cf.~\cite[Thm. 3.9]{Chavel-1993}).
\end{proof}  

Proposition \ref{LiYau} can be extended to the perturbed operator. To
do so we will need the Feynman-Kac formula:
\[ 
e^{-t H_\omega}f(x) = \Erw(\exp(-\int_0^t V_\omega(X_s)ds) f(X_t)),
\]
where $\Erw$ denotes the expectation with respect to the Brownian
motion $X_t$ starting in $x$. This formula on
stochastically complete manifolds is presented, e.g., in
\cite[Thm. IX.7A]{Elworthy-82} for bounded continuous
potentials. Using semigroup and integral convergence theorems in the
same spirit as in the proof of \cite[Thm. X.68]{ReedS-75}, the
validity of this formula can be extended to nonnegative, locally $L^1$
potentials. 

\begin{coro} \label{LiYausuper}
For arbitrary open $D$ and $t > 0$ the following holds
\begin{equation}\label{boundsdomain}  
0\leq \kD(t,x,y) \leq \ko (t,x,y) \leq C_t \exp (-\alpha_t
d^2_\omega(x,y)),
\end{equation}  
for almost all $x,y \in D$. The constants are as in the previous
proposition. In particular, the integral estimate in \ref{LiYau}(ii)
holds also for the perturbed operator, for almost all $x \in X$.
\end{coro}

\begin{proof}
Using the Feynman-Kac formula, we obtain
\begin{multline*}
0 \le \int_X k_\omega(t,x,y)f(y)dy = \Erw(\exp(-\int_0^t V_\omega(X_s)ds)
f(X_t)) \\
\le \Erw(f(X_t)) = \int_X k_{\Delta_\omega}(t,x,y)f(y)dy, 
\end{multline*}
for all nonnegative $f \in L^2(X,\vol_\omega)$. Proposition \ref{LiYau} 
implies
\[ 
0 \le k_\omega(t,x,y) \le k_{\Delta_\omega}(t,x,y) \le C_t \exp 
(-\alpha_t d^2_\omega(x,y)), 
\] 
for almost all $x, y \in X$. The inequality for the Dirichlet operator
follows by so called \emph{domain monotonicity}, see
e.g.~\cite[Thm.~2.1.6]{Davies-1989}. The proof of this theorem carries
directly over from $\RR^n$ to manifolds.
\end{proof}

Finally, note that, for $\vol_\omega (D) <\infty$, the estimate $0 \le
k_\omega^D(t,x,y) \le C_t$ implies that $\ExpHD$ is a
Hilbert-Schmidt operator and thus $\HD$ has purely discrete spectrum
by the spectral mapping theorem.

\section{The principle of not feeling the boundary}  
\label{PNFB}  

In this section we show that the semigroups associated to our random
operators satisfy a principle of not feeling the boundary.
\medskip  
  
Let $D$ be a open set on the manifold $X$. One expects the difference
between the Dirichlet heat kernel $k_{H^D} (t,x,y)$ and $k_{H}
(t,x,y)$ to be small as long as $t > 0$ is small and $x$ and $y$ stays
away from the boundary of $D$. This phenomenon is called {\em
principle of not feeling the boundary}. To treat it rigorously, we
introduce the notion of thickened boundary.
%Henceforth, we refer to a
%{\em region} as a bounded open set $D$, consisting only of finitely many
%connected components. 
For $ h >0$, let $\partial_{h} D:= \{x \in X| \, d_0(x,\partial D) \le
h \}$ and $D_h$ be the interior of the set $D \setminus \partial_{h}
D$. Note that (M2) of Definition \ref{randommetric}
implies the following inequalities 
\[
C_g^{-1} d_0(x,y) \le d_\omega(x,y) \le C_g d_0(x,y).
\] 
The main result of this section is the following:
  
\begin{theorem} \label{NFTB} For all $t, \epsilon>0$, there exists an 
$h=h(t,\epsilon)>0$ such that for every open set $D\subset X$ and
all $\omega \in \Omega$, we have
\[ 0\leq \ko (t,x,y) - \kD (t,x,y)\leq \epsilon, \]
for almost all $x, y\in {D_h}$.
\end{theorem}  
  
The \textit{Proof of the Theorem} follows from the next two propositions. More precisely, in view of the next proposition, it is enough to prove the theorem for vanishing potential. This, however, is accomplished in Proposition \ref{freierfall}.

Let $\tau_x^D$ denote the first exit
time from $D$ for Brownian motion starting in $x$.

\begin{prop} The following statements are equivalent. 

(i) For all $t, \epsilon>0$, there exists an $h=h(t,\epsilon)>0$
such that for every open set $D\subset X$ and all $\omega \in
\Omega$, we have
\[ 0\leq k_{\Delta_\omega}(t,x,y) - k_{\Delta_\omega^D} (t,x,y)\leq  
\epsilon, \] for almost all $x, y \in {D_h}$. 
 
(ii) For all $t, \epsilon>0$, there exists an $h=h(t,\epsilon)>0$ such
that for every open set $D\subset X$ and all $\omega \in \Omega$
\[ \Erw (\chi_{D_h} (X_t) \chi_{ \tau_x^D < t} )\leq \epsilon, \]
for almost every $x\in D_h$. 

(iii) For every random operator $\{ H_\omega \}$ and for all $t,
\delta>0$, there exists an $r=r(t,\epsilon,H)>0$ such that for every
open set $D\subset X$ and all $\omega \in \Omega$, we have
\[ 0\leq \ko (t,x,y) - \kD (t,x,y)\leq \delta, \] 
for almost all $x, y \in {D_r}$.
\end{prop}  

\begin{proof} 
(i)$\Longrightarrow$(ii). By the Feynman-Kac formula for the
unperturbed $\Delta_\omega$-operator, we have
\[  
\Erw( \chi_{D_h} (X_t) \chi_{\{\tau_x^D <t\}} ) = \int
[k_{\Delta_\omega} (t,x,y) - k_{\Delta_\omega^D}(t,x,y)] \,
\chi_{D_h}(y) \, dy
\]  
which can be bounded using the H\"older inequality and domain
monotonicity by
\begin{multline*}  
\esssup_{y\in D_h} |k_{\Delta_\omega} (t,x,y) -
k_{\Delta_\omega^D}(t,x,y))|^{\frac{1}{2}} \int (k_{\Delta_\omega}
(t,x,y) - k_{\Delta_\omega^D}(t,x,y))^{\frac{1}{2}} \chi_{D_h} (y) dy
\\ \leq \esssup_{y\in D_h} |k_{\Delta_\omega} (t,x,y) -
k_{\Delta_\omega^D}(t,x,y)|^{\frac{1}{2}} \int k_{\Delta_\omega}
(t,x,y)^{\frac{1}{2}} \chi_{D_h} (y) dy
\end{multline*}  
The first term can be seen to be small by (i) for almost every $x \in
D_h$, and the second term is bounded by $B_{t,1/2}$, due to Proposition 
\ref{LiYau} (ii).
\medskip  
  
(ii)$\Longrightarrow$(iii). We have to show that 
\[  
\chi_{D_h}(\ExpH -\ExpHD)\chi_{D_h}\colon L^1 (D_h, \vol_\omega) \rightarrow L^\infty( D_h, \vol_\omega)
\]  
is arbitrarily small for $h$ large enough (independently of $\omega$
and $D$). Let $R := \exp(-\frac{t}{2} H_\omega) - \exp(-\frac{t}{2} H_\omega^D)$. Note that
\[ \Vert e^{-\frac{t}{2}H_\omega^D} \Vert_{1\to 2} \le \sqrt{B_{t/2,2}}, \]
by Corollary \ref{LiYausuper}. Thus, since
\[
\chi_{D_h}(e^{-t H_\omega} -e^{-t H_\omega^D})\chi_{D_h} = \chi_{D_h}
e^{-\frac{t}{2} H_\omega} R \chi_{D_h} + \chi_{D_h} R e^{-\frac{t}{2}
H_\omega^D}\chi_{D_h}
\]
and by duality 
\[
\| R \chi_{D_h}  \|_{1\to 2}
=
\| \chi_{D_h} R  \|_{2\to \infty}
\]
it suffices to show that
\[  
\chi_{D_h}(\ExpH -\ExpHD) \chi_{D_h}
 \colon L^2 (D_h, \vol_\omega) \rightarrow L^\infty( D_h, \vol_\omega)
\]  
is arbitrarily small for $h$ large enough (independently of $\omega$
and $D$). Using the Feynman-Kac Formula for the perturbed operator, we
obtain
\begin{eqnarray*}  
\lefteqn{(\ExpH -\ExpHD) f (x)} \\ 
& =& \Erw (\exp(-\int_0^t V_\omega(X_s)\,ds) \chi_{D_h} (X_t) f(X_t) \chi_{\{
\tau_D^x <t\}} ) \\ 
& \leq & \Erw (f(X_t)^2)^{\frac{1}{2}} \Erw( \chi_{D_h} (X_t)
\chi_{\{\tau_D^x <t\}} )^{\frac{1}{2}} \\ 
&\leq&(\int_X \ko (t,x,y) |f(y)|^2 dy)^{\frac{1}{2}} \Erw( \chi_{D_h}
(X_t) \chi_{\{\tau_D^x <t\}} )^{\frac{1}{2}} \\ 
&\leq& {C_t}^{1/2} \|f\|_{2} \ \Erw( \chi_{D_h} (X_t) \chi_{\{\tau_D^x <t\}}
)^{\frac{1}{2}}.
\end{eqnarray*}  
%  \begin{multline*}  
%  (\ExpH -\ExpHD) f (x) 
%    =  \Erw (\exp(-\int_0^t V_\omega(X_s)\,ds) \chi_{D_r} (X_t) f(X_t) \chi_{\{
%  \tau_D^x <t\}} ) \\ 
%    \leq   
%    \Erw (f(X_t)^2)^{\frac{1}{2}} \Erw( \chi_{D_r} (X_t)
%  \chi_{\{\tau_D^x <t\}} )^{\frac{1}{2}} 
%  \\ \leq 
%  (\int_X \ko (t,x,y) |f(y)|^2 dy)^{\frac{1}{2}} \Erw( \chi_{D_r}
%  (X_t) \chi_{\{\tau_D^x <t\}} )^{\frac{1}{2}}  
%   \leq  
%   C_t \|f\|_{2} \ \Erw( \chi_{D_r} (X_t) \chi_{\{\tau_D^x <t\}})^{\frac{1}{2}}.
%  \end{multline*}
The proof is finished by invoking (ii).  
\medskip  
  
(iii)$\Longrightarrow$(i). This is immediate by choosing $V_\omega\equiv 0$.  
\end{proof}

The following lemma is an adaptation of Proposition 1.1 in
\cite[Chp.~6]{Taylor-96a}. It is useful in our proof of ``not feeling
the boundary'':

\begin{lemma}[Maximum principle for heat equation with nonnegative 
potential] Let $D \subset X$ be open with compact closure, $V \ge 0$,
and $u \in C ( [0, T[ \times \overline{D}) \, \cap \, C^2 ( ]0, T[
\times D)$ be a solution of the heat equation
$\frac{\partial}{\partial t} u + (\Delta + V)u = 0$ on $ ]0, T[ \times
D$ with nonnegative supremum $s = \sup\{ u(t,x) \mid (t,x) \in [0,T[
\times \overline{D} \}$. Then,
$$
s = \max \l \{ \max_{x \in \overline{D}} u (0,x) , \sup_{[0, 
T[ \times \partial D} u (t,x) \r \}.
$$
\end{lemma}

\begin{proof}  
It suffices to prove that, for any $c \ge 0$, the assumption
\begin{equation} \label{asum}
u  <c \text{ on } (\{0\} \times \overline{D}) \cup ([0, T[ \times \partial D)
\end{equation}
implies $u \le c$ on $[0, T[ \times D$. To this aim we introduce the
auxiliary function $u_\delta(t,x) = u(t,x) - \delta t$, $\delta > 0$,
and show that \eqref{asum} implies $u_\delta(t_0,x_0) < c$ on $[0,T[
\times D$.

Assume that the conclusion is wrong. Then there exists $(t_0,x_0) \in
]0,T[ \times D$ such that $u_\delta(t_0,x_0) \ge c$. By continuity the
function $f(t) := \max_{x \in \overline{D}} u_\delta (t,x)$ is well
defined and $t_1 := \min_{t \ge 0} \{ t | \, f(t) = c \}$
exists. By \eqref{asum}, we have $0 < t_1 \le t_0$, and there exists an $x_1
\in D$ such that $u_\delta (t_1, x_1) = c$.
  
On the one hand we have $ \frac{\partial u_\delta}{\partial t} ( t_1,
x_1) \ge 0$ and, on the other, since $u_\delta(t_1,\cdot)$ has a
global maximum at $x_1$: $(\Delta_x u_\delta)(t_1,x_1) \ge 0$. Evaluating at $(t_1,x_1)$ 
yields the desired contradiction:
\begin{equation*}
0 \le \frac{\partial u_\delta}{\partial t}   = \frac{\partial u}{\partial t}  -\delta
= -\Delta u -Vu - \delta \le  - \delta < 0
\end{equation*}
\end{proof}
  
We now prove the principle of not feeling the boundary for the free
Laplacian using an idea of H.~Weyl (cf. \cite[Lemma 3.5]{Dodziuk-1981}
for a Euclidean version).
  
\begin{prop} \label{freierfall}  
For any fixed $t, \epsilon>0$, there exists an $h=h(t,\epsilon)>0$ such
that for every open set $D\subset X$ and all $\omega \in \Omega$
\[  
0\leq k_{\Delta_\omega}(t,x,y) - k_{\Delta_\omega^D} (t,x,y)\leq
\epsilon,
\]  
for all $x\in D, y \in {D_h}$.
\end{prop}  

\begin{proof} We prove that the proposition is true for any $h > 0$ 
satisfying
\[
C_t \exp(-\alpha_t C_g^{-2} (h/2)^2) \le \epsilon.
\]
Let $\omega \in \Omega$ be fixed, and $f_\delta\in C_0^\infty (B_\delta
(y))$, with $0 < \delta < h/2$, be a nonnegative approximation of the
$\delta_y$-distribution at $y \in D_h$. Here, $B_\delta (y)$ denotes
the open $d_\omega$-ball around $y$ with radius $\delta$. Denote by
$k(t,x,y)=k_{\Delta_\omega}(t,x,y)$ the heat kernel of the semigroup
$e^{-t\Delta_\omega}$ and set
\[ 
u_1 (t,x):= \int_X k(t,x,z) f_\delta(z) d\vol_\omega(z) = \int_D
k(t,x,z) f_\delta(z) d\vol_\omega(z).
\]
Moreover, let $k^D (t,x,y)= k_{\Delta_\omega^D}(t,x,y)$ be the heat
kernel of the semigroup $e^{-t\Delta_\omega^D}$ on $D$
with Dirichlet data on the boundary $\partial D$, and set 
\[
u_2 (t,x):= \int_D k^D(t,x,z) f_\delta(z) d\vol_\omega(z).
\] 
The difference $u_1(t,x)-u_2(t,x)$ solves the differential equation
$\left( \frac{\partial }{\partial t} + \Delta_\omega \right) u = 0$
and satisfies the initial condition $u_1(0,x)-u_2(0,x)=
f_\delta(x)-f_\delta(x)=0$ for all $x \in D$. Now, by domain
monotonicity we know $k(t,x,z)-k^D(t,x,z) \ge 0$, thus 
\bea
u_1(t,x)-u_2(t,x)= \int_D (k(t,x,z) -k^D (t,x,z)) f_\delta(z)
d\vol_\omega(z) \ge 0
\eea 
for all $t >0$ and $x \in D$. The application of the maximum principle yields
\begin{equation} \label{Max-am-Werk}  
u_1(t,x)-u_2(t,x) \le \max_{]0,t] \times \partial D} \l\{
u_1(s,w)-u_2(s,w) \r\}.
\end{equation}
The expression on the right hand side can be further estimated as:
\begin{align*}  
u_1(s,w)-u_2(s,w) \le \int_D k(s,w,z) f_\delta(z) d\vol_\omega(z) 
= \int_{D_{h/2}} k(s,w,z) f_\delta(z) d\vol_\omega(z).
\end{align*}  
Since $w \in \partial D$ and $z \in D_{h/2}$, we conclude with \eqref{kdelest}
in Proposition \ref{LiYau}:
\[
\int_{D_{h/2}} k(s,w,z) f_\delta(z) d\vol_\omega(z) \le C_t
\exp(-\alpha_t C_g^{-2} (h/2)^2) \le \epsilon.
\]
Taking the limit $\delta \to 0$, proves the proposition.
\end{proof}

\section{Construction of the IDS by an exhaustion procedure}\label{construct}  

Using the strategy of \cite{PeyerimhoffV-2002}, we show that the IDS,
defined in \eqref{absIDS}, coincides with the limit of an exhaustion
procedure, for almost all $\omega \in \Omega$. This proves the
selfaveraging property of the IDS stated in Theorem \ref{selfaverIDS}.
  
We first introduce the notion of an admissible sequence of subsets of
$X$.  As explained in \cite[Section 3]{AdachiS-1993}, let $\calF
\subset X$ be a polyhedral fundamental domain of the group
$\Gamma$. Any finite subset $I \subset \Gamma$ defines a corresponding
set
\[ 
\phi(I) := {\rm int}\bigg ( \bigcup_{\gamma \in I} \gamma \overline{\calF}
\bigg ) \subset X.
\]
Now, admissible sequences are defined via tempered F{\o}lner sequences:

\begin{definition} \label{admseq}  
(a) A sequence $\{I_j\}_j$ of finite subsets in $\Gamma$ is called a {\em
  F{\o}lner sequence} if $ \lim_{j \to \infty} \frac{\vert I_j \Delta
  I_j \gamma \vert} {\vert I_j\vert} = 0$ for all $\gamma \in
  \Gamma$.\\
(b) A F{\o}lner sequence $\{I_j\}_j$ is called a {\em tempered
  F{\o}lner sequence} if it is monotonously increasing and satisfies
  $\sup_{j \in \NN} \frac{\vert I_{j+1} I_j^{-1} \vert}{\vert I_{j+1}
  \vert} < \infty.$\\
(c) A sequence $\{D^j\}_j$ of subsets of $X$ is called {\em
  admissible} if there exists a tempered F{\o}lner sequence
  $\{I_j\}_j$ in $\Gamma$ with $D^j = \phi (I_j)$, $j\in \NN$.
\end{definition}  
  
By Lemma 2.4 in \cite{PeyerimhoffV-2002}, an admissible sequence
satisfies the isoperimetric property
\begin{equation} \label{isop}  
\lim_{j \to \infty} \frac{\vol_0(\partial_d D^j)}{\vol_0(D^j)} = 0, \quad
  \text{for all $d > 0$}.
\end{equation} 
Existence of a F{\o}lner sequence is a geometrical description of
amenability of the group $\Gamma$. The notion of `tempered F{\o}lner
sequence'' is due to A. Shulman \cite{Shulman-1988} and used by
Lindenstrauss in the proof of the following pointwise ergodic theorem
\cite{Lindenstrauss-2001}.
    
\begin{theorem} \label{ergthm}  

(a) Every F{\o}lner sequence has a tempered subsequence. In
particular, every amenable group admits a tempered F{\o}lner sequence.
  
(b) Let $\Gamma$ be an amenable discrete group and
$(\Omega,\calA,\PP)$ be a probability space. Assume that $\Gamma$ acts
{\em ergodically} on $\Omega$ by measure preserving transformations
$\{ T_\gamma \}_\gamma$. Let $\{ I_j \}_j $ be a tempered F{\o}lner
sequence. Then we have, for every $f \in L^1(\Omega)$ 
\begin{equation} \label{aver} 
\lim_{j \to \infty} \frac{1}{\vert I_j \vert} \sum_{\gamma \in
I_j^{-1}} f(T_\gamma \omega) = \EE(f) 
\end{equation} 
in almost-sure and $L^1$-topology.
\end{theorem}  
  
The results of the last two sections are used to prove the following
heat kernel lemma:  
  
\begin{lemma} \label{heatkernellemma}
Let $\{ D^j \}$, $j\in \NN$, be an admissible sequence and let
$\{H_\omega\}_\omega$ be a random operator. Then the following
holds.
\\ 
(a) $\sup_{\omega\in \Omega} \ \vol_\omega(D^j)^{-1}
\left|  \tr (\chi_{D^j} \ExpH) - \tr( \ExpHDk)\right| 
\rightarrow 0, \ n\to
\infty.$
\\ 
(b) There exists a constant $C>0$ with $\tr (\chi_{\calF}
e^{- t H_\omega}) \leq C$ for all $\omega\in \Omega$.
\\ 
(c) The map
$\omega\mapsto \tr (\chi_{\calF} e^{- t H_\omega})$ is measurable.
\end{lemma}

\begin{proof} 
(a) By $\exp(-t H) = \exp(-\frac{t}{2} H) \exp(\frac{t}{2} H)$ for
arbitrary $H\geq 0$ and standard calculations for integral kernels, we have
\begin{equation}
 \label{HGHS} 
\tr (\chi_{D^j} e^{-t H_\omega})=\int_{D^j} \int_{D^j} \ko
(t/2,x,y)^2 d\vol_\omega(x) d\vol_\omega(y)
\end{equation} 
and
\begin{equation}
 \label{HGHSr} 
\tr(e^{-t H_\omega^{j}}) = \int_{D^j} \int_{D^j} k_{H_\omega^{j}}
(t/2,x,y)^2 d\vol_\omega(x) d\vol_\omega(y).
\end{equation} 
We express the difference of \eqref{HGHS} and \eqref{HGHSr} using
$\ko^2 - \kDk^2=[\ko - \kDk][\ko + \kDk]$ and use the following
decomposition of the integration domain:
\begin{multline*}
\int_{D^j} \int_{D^j} [\ko(t/2,x,y) - \kDk(t/2,x,y)] \, [\ko(t/2,x,y)
 + \kDk(t/2,x,y)] \, d\vol_\omega(x,y) \\ = {\int_{\partial_h D^j}
 \int_{\partial_h D^j} \dots} + {\int_{D_h^j} \int_{\partial_h D^j}
 \dots} + {\int_{\partial_h D^j} \int_{D_h^j} \dots} + \int_{D_h^j}
 \int_{D_h^j} \dots
\end{multline*}
Each of the first three terms can be bounded by $2 C_{t/2} B_{t/2} \vol_\omega
(\partial_h D^j)$ by inferring the following consequences of Section \ref{HK}:
\begin{equation*} 
0\leq \kDk (t/2,x,y)\leq \ko(t/2,x,y)\leq C_{t/2}  \mbox{ and }  \int \ko
(t/2,x,y)\,d\vol_\omega(y) \leq B_{t/2}
\end{equation*}  
for almost every $x,y\in X$. As for the last term, we fix $\epsilon>0$
and choose $h=h(t/2,\epsilon)$ according to Theorem \ref{NFTB} and
obtain the bound $2 \epsilon B_{t/2} \vol_\omega (D^j)$.  By
\eqref{rhoest}, we conclude that the sequence $D^j$ satisfies the
isoperimetric property
\[ 
\lim_{j \to \infty} \frac{\vol_\omega(\partial_d D^j)}{\vol_\omega(D^j)} = 0,
  \quad \text{for all $d > 0$},
\]
for the metric $g_\omega$, as well. This shows part (a).  

Now, (b) follows from \eqref{rhoest}, Proposition \ref{LiYau} (ii) and
the analog of \eqref{HGHS} for $\chi_{\calF}$, while (c) follows from
measurability of $\omega \mapsto e^{- tH_\omega}$, after choosing a
suitable orthonormal basis according to Appendix A of
\cite{LenzPV-2002}.
\end{proof}  
  
Finally, we present the proof of our main result:
  
\begin{proof}[Proof of Theorem \ref{selfaverIDS}] 
A criterion of Pastur and \v Subin \cite{Pastur-1971,Shubin-1979} establishes the convergence of the normalized eigenvalue counting functions $N_\omega^j$ to a selfaveraging limit, if their Laplace transforms $\calL_\omega^j(\cdot)$ converge. 
To apply this criterion (cf.~\cite{PeyerimhoffV-2002, LenzPV-2002}, as well) we note that the random operator $\{H_\omega\}_\omega$ is non-negative and by \ref{heatkernellemma} (b) the $\calL_\omega^j(t)$ are bounded by a constant depending only on $t$. So it remains to show:
\[  
\lim_{j \to\infty} \calL_\omega^j(t) :=
\lim_{j \to\infty}  \int_\RR e^{-t\lambda} dN_\omega^j (\lambda) =  
\int_\RR e^{-t\lambda } dN_H(\lambda) 
\]  
for all $t > 0$, in $L^1$ and $\PP$ almost sure-sense. This is done by
applying Lindenstrauss' ergodic theorem (Theorem \ref{ergthm}). We
introduce the equivalence relation $a_j\stackrel{j\to \infty}{\sim}
b_j$ for two arbitrary sequences $a_j(\omega),b_j(\omega), j \in \NN$,
satisfying $a_j - b_j \to 0$, as $j \to \infty$, in $L^1$ and
$\PP$-almost surely. By definition we have
$$
\calL_\omega^j(t) = \vol_\omega(D^j)^{-1} 
\tr(e^{-tH_\omega^j}).
$$
Using the previous lemma, equivariance and Theorem \ref{ergthm}, we
derive
\begin{eqnarray*}  
\vert I_j \vert^{-1} \tr(e^{-t H_\omega^j}) &\stackrel{j\to
\infty}{\sim}& \vert I_j \vert^{-1} \tr (\chi_{D^j} e^{-t H_\omega}) \ = \ 
\vert I_j\vert^{-1} \sum_{\gamma \in I_j} \tr( \chi_{\gamma \calF}
e^{-t H_\omega})\\
&=& \vert I_j\vert^{-1} \sum_{\gamma \in I_j^{-1}} \tr(\chi_{\calF}
e^{-t H_{\gamma \omega}})
\ \stackrel{j\to \infty}{\sim} \ \EE ( \tr (\chi_{\calF} e^{-t H_\bullet})).
\end{eqnarray*}  
Similarly, we infer  
\begin{equation*}  
\vert I_j\vert^{-1} \vol_\omega(D^j) = \vert I_j\vert^{-1}
\sum_{\gamma \in I_j} \vol_\omega(\gamma \calF) = \vert I_j\vert^{-1}
\sum_{\gamma \in I_j^{-1}} \vol_{\gamma \omega}(\calF)
\stackrel{j\to \infty}{\sim} \EE \l\{\vol_\bullet( \calF)\r\} .
\end{equation*}  
Putting this together, and noting that, by \eqref{rhoest},
\[ 
C_g^{-n/2}\vol_0(\calF) \le |I|^{-1} \vol_\omega(\phi(I)) \le
C_g^{n/2}\vol_0(\calF),
\] 
for all finite sets $I \subset \Gamma$, we obtain
\begin{multline*}
\calL_\omega^j(t) = \vol_\omega(D^j)^{-1}
\tr(e^{-tH_\omega^j}) = \frac{\vert I_j\vert^{-1} \tr
(e^{-tH_\omega^j})}{\vert I_j\vert^{-1} \vol_\omega(D^j)} 
\stackrel{j\to \infty}{\sim} 
\frac{\EE\l\{ \tr(\chi_\calF e^{-tH_\bullet})\r\}}{\EE \l\{\vol_\bullet(
\calF)\r\}}.
\end{multline*}
By \eqref{absIDS},
\[ 
\frac{\EE\l\{ \tr(\chi_\calF e^{-tH_\bullet})\r\}}{\EE \l\{\vol_\bullet(
\calF)\r\}} = \int_\RR e^{-t \lambda} dN_H(\lambda).
\]  
This finishes the proof.  
\end{proof}  
  
{\small \textbf{ Acknowledgements:} It is a pleasure to thank B.~Franke,
D.~Hundertmark, L.~Karp, W.~Kirsch, O.~Post and P.~Stollmann for stimulating
discussions.  This work was supported in part by the DFG through SFB
237 ``Unordnung und gro{\ss}e Fluktuationen'' and the
Schwer\-punkt\-programm "Interagierende stochastische Systeme von
hoher Komplexit\"at".}
  
\def\cprime{$'$}

\end{document}